\documentclass[12pt]{article}

\usepackage[english]{babel}
\usepackage{comment}
\usepackage[letterpaper,top=2cm,bottom=2cm,left=3cm,right=3cm,marginparwidth=1.75cm]{geometry}

\usepackage{amsmath}
\usepackage{graphicx,natbib}
\usepackage[colorlinks=true, allcolors=blue]{hyperref}
\usepackage{multirow}
\usepackage{graphics,booktabs}
\usepackage{amsthm,amssymb,authblk}

\usepackage{amsmath, amsthm, amssymb,pdflscape,bm,rotating,color,colortbl}
\usepackage{graphicx}
\usepackage{tabularx}
\usepackage{psfrag}
\usepackage{fancyhdr}
\usepackage{lscape,pdflscape}
\usepackage{times,url}

\usepackage{algorithm,algpseudocode}

\title{On the estimation of complex statistics combining different surveys}
\author{Vasilis Chasiotis}
\author{Dimitris Karlis}

\affil{\small Department of Statistics, Athens University of Economics and Business,  Greece}

\date{}

\begin{document}
\maketitle

\begin{abstract}
The importance of exploring a potential integration among surveys has been acknowledged in order to enhance effectiveness and minimize expenses. In this work, we employ the alignment method to combine information from two different surveys for the estimation of complex statistics. The derivation of the alignment weights poses challenges in case of complex statistics due to their non-linear form. To overcome this, we propose to use a linearized variable  associated with the complex statistic under consideration. Linearized variables have been widely used to derive variance estimates, thus allowing for the estimation of the variance of the combined complex statistics estimates. Simulations conducted show the effectiveness of the proposed approach, resulting to the reduction of the variance of the combined complex statistics estimates. Also, in some cases, the usage of the alignment weights derived using the linearized variable associated with a complex statistic, could result in a further reduction of the variance of the combined estimates.
\end{abstract}


\section{Introduction}
It has been recognized the need to investigate the possibility to integrate different surveys to improve efficiency and reduce costs \citep{hidiroglou2001,merkouris2010,wu2004,ybarra2008}. The combination of samples from different surveys, or even different samples of the same survey, can result in a more representative sample of the target population allowing a more precise estimation of parameters. This can be particularly beneficial in case that individual surveys may have limitations.

A method of pooling information from different surveys is through alignment, that is when two separate surveys have been conducted within the same population and have common variables, then it may be preferable the modification of the weights of each survey so as to ensure that they result in equal estimates for these common variables. \cite{zieschang1990} introduced a methodology of composite estimation involving an extended regression constraint system. \cite{renssen1997} extended the idea of \cite{zieschang1990} and proposed a class of composite regression estimators. In the two-step method proposed by \cite{renssen1997}, separate initial estimates are first calculated for the common variables within each sample. These individual estimates are then combined to derive a common estimate, incorporating the information of both samples. Then each sample treats the common variable as an auxiliary variable, and the common estimate as the known population total. Using this approach, new estimates for all variables are calculated. \cite{merkouris2004} proposed a  one-step method, in which one would need all micro data from both samples, and then the alignment weights are calculated on the basis of a regression (equivalently calibration) procedure that is implemented on the combined data. In particular, the proposed method by \cite{merkouris2004} is a modification of Zieschang's method that accounts for the differential in effective sample size between two surveys. This modification retains the practical advantages of Zieschang's method as well as improves the efficiency of the derived composite regression estimators. \cite{kim2012} proposed a model-assisted method to combine data from two independent surveys. In this method, a large sample from one survey gathers solely auxiliary information, and a considerably smaller sample from another survey provides information on both the variables of interest and the auxiliary variables. More recently, \cite{berger2020} developed a design-based empirical likelihood approach for aligning information from multiple surveys and estimating complex parameters defined by estimating equations. \cite{yang2020} provided a comprehensive overview of data integration in survey sampling.

Complex statistics refer to quantitative measures that involve intricate mathematical expressions, such as non-linear relationships among multiple variables. Examples of such statistics, among others, include the median, the quintile share ratio S80/S20, the Gini coefficient, the at-risk-of-poverty rate, and the relative median at-risk-of-poverty gap, all of them useful in social statistics. The alignment method of \cite{merkouris2004} is developed for totals, and as such, its application to complex statistics is not straightforward. Thus, the derivation of the alignment weights are rather a cumbersome non-linear optimization problem. This aspect highlights the challenges when
dealing with complex statistics. In this paper,  in order to overcome this obstacle, we propose to make usage of the linearized variable associated with the complex statistic as it is described in \cite{deville1999}. As of now, the aligment method of \cite{merkouris2004} can only be applied using totals. Our proposed approach offers the opportunity to obtain alignment weights considering complex statistics, thus making the pooling of information from two different surveys achievable. Also, the method of \cite{deville1999} proposed for variance estimation; the variance of the linearized variable under a sampling design provides an approximation of the variance of the complex statistic considered. As a consequence, an approximation of the variance for our proposed combination of the complex statistics estimates can be derived. Furthermore, the usage of alignment weights obtained by complex statistics could result in a further reduction of the variance of the combined estimates.

\begin{comment}
{\color{red} There are some packages in {\texttt R} developed for variance estimation in complex surveys. Package {\texttt laeken} \citep{alfons2013} developed for most of the complex statistics, and is an object-oriented toolkit for the estimation of indicators from complex survey samples via standard or robust methods. In particular the most widely used social exclusion and poverty indicators are implemented in the package. A general calibrated bootstrap method to estimate the variance of indicators for common survey designs is included as well. The package {\texttt ineq} \cite{zeileis2014} calculates various inequality, concentration and poverty measures. Also, \cite{kainu2016} developed a package for merging EU-silc cross-sectional and longitudinal raw ``.csv" datasets, called \texttt{r.eusilc}. Moreover, the package \texttt{vardpoor}, see \cite{breidacks2015}, calculates the linearized variables for most of the
complex statistics.}
\end{comment}

The remaining of this paper proceeds as follows. Section \ref{background} provides a description on how variance is estimated for all complex statistics according to the technique
proposed by \cite{deville1999}, as well as some technical details about the alignment method of \cite{merkouris2004}. In Section \ref{complex} the linearized variables for specific complex statistics addressed in this paper are provided. In Section \ref{our_approach} the proposed approach of estimating complex statistics pooling information from different surveys through alignment is described. The dataset to be used for the implementation as well as the simulation results of the proposed approach are provided in Section \ref{results}, while concluding remarks can be found in Section \ref{concl}.

\section{Background}\label{background}
In this section we provide some background information related to material that we will use later on. In particular we describe the idea of a linearized variable used for the variance estimation method for complex statistics proposed by \cite{deville1999}.
This quantity will be the basis of our derivations. Also, we describe the alignment method of \cite{merkouris2004} that is applied in Section \ref{our_approach} for pooling information from two different surveys for the estimation of complex statistics.

\subsection{Variance estimation for complex statistics}
Linearization is a technique that allows the variance estimation of complex statistics, by approximating a complex estimator by a simpler, linear estimator. The idea is to find a linearized variable that relates to the complex statistic under consideration.  This simplifies the calculation of the variance of the estimator, since the variance of the linear approximation can be used as an approximation of the variance of the non-linear indicator considered. 

The proposed variance estimation is based on the linearization method proposed by \cite{deville1999}, for which the basis is the influence function. We consider a finite population $U$ of $N$ units that are denoted by $u_i$, $i=1,\ldots,k,\ldots,N$. The underlying concept of the linearization method proposed by \cite{deville1999} is to examine the impact of unit $k$ on the population parameter of interest by introducing an infinitesimal variation in the weight assigned to this unit. Let $M$ be a measure that allocates a unit mass to all $k \in U$. A population parameter $\theta$ can be written as a functional $T(M)$, and so the influence function of $T$ is defined as  
\[
z_k = I[T(M)]_k = \lim_{t \to 0} \frac{T(M + t \, d_k) - T(M)}{t},
\]
where $d_k$ denotes the Dirac measure for the $k$-th unit and $z_k$ denotes the linearized variable 
for the $k$-th unit.

Let $\widehat{\theta}$ be an estimator of $\theta$ and $w_k$ be the weight associated to the $k$-th unit. It is shown \citep{deville1999}, under asymptotic conditions, that:
\begin{equation}\label{VAR}
Var\left(\sum_{k \in S} z_k w_k\right) \approx Var \left(\widehat{\theta}\right).
\end{equation}
However, the $z_k$'s are unknown since they depend on unavailable information regarding the population. Let $S$ be a random sample from $U$ according to a sampling design $p(s)=\text{P}(S=s)$, for all $s\in S$, and $\widehat{M}$ be the measure allocating a mass $w_k$ to all $k \in S$. The $z_k$'s can be estimated from $S$ using the plug-in estimator $\widehat{z_k} = I[T(\widehat{M})]_k$. Therefore, under any sampling design, the variance of a complex statistic $\widehat{\theta}$ can be estimated.

For general designs the variance can be estimated using the first
and second-order inclusion probabilities. Let $\pi_k$ be the first-order inclusion probability for the
$k$-th unit and $\pi_{k\ell}$ be the second-order probabilities of
inclusion in $S$ of both the $k$-th and the $\ell$-th units. Defining
\[
\tilde \Delta_{k \ell} = \frac{\pi_{k\ell}- \pi_k
\pi_\ell}{\pi_{k\ell}},
\]

the variance of $\widehat \theta$ can be estimated as
\begin{equation}
{\widehat{Var}}(\widehat \theta) = \sum\limits_{k \in S} \sum\limits_{\ell
\in S} \tilde \Delta_{k \ell}  \frac{\widehat{z_k} \widehat{z_\ell}}{ \pi_k \pi_\ell}.
\label{variance}
\end{equation}

The aforementioned approach provides the framework for obtaining alignment weights considering complex statistics and deriving an approximation of the variance for the combination of the complex statistics estimates, as detailed in Section \ref{our_approach}.

\subsection{Pooling information from different surveys}\label{back2}

Pooling information from two or more independent surveys is a problem frequently encountered in survey sampling.
Appropriate pooling can lead to reducing the cost and improve the efficiency of sample surveys and as such it has received considerable interest. 

The pooling information can be derived either from independent surveys of the same target population or from two independent samples of the same survey. The idea of alignment suggests that if two sample surveys have some variables in common, then it may
be attractive for both surveys to give virtually the same estimates for the population totals of these common variables. We use the term common variables for those variables observed in both surveys but for which the corresponding population totals are unknown. Typical examples of such variables include household
characteristics such as size or composition of the household. For each survey, it is generally required that known population totals of auxiliary (control) variables be reproduced by weighting-type estimators. Examples of such control variables are sex, age, marital status, or region. For these variables, the population totals are known and thus we can force their
estimates from the sample to fit the known frequencies.

The requirement of reproducing these counts is always fulfilled if one takes these control variables as regressors in the general regression estimator. For the common variables, the population totals are estimated by pooling both surveys, and then simultaneously using these common variables as additional regressors. The implicitly defined weights of this adjusted
regression estimator reproduce the known population totals (or frequencies) of the control variables, as well as the estimates of the population totals of the common variables. These weights are called the alignment weights. The weights of the adjusted regression estimator are called reproductive with respect to the control variables and consistent with respect to the common variables. 

The usage of alignment weights derived from simpler statistics, like totals, have some interesting properties \citep{zieschang1990} including  optimality in the sense that they match known estimators, improved interpretation, rather easy derivation and they also correct for any problems in each
sample.

In this paper, we follow the alignment method of \cite{merkouris2004}.  We assume that the two samples $S_1 $ and $S_2 $ of sizes $n_1$ and $n_2$, respectively, have
observations  ${\rm {\bf Y}}_1 $ and ${\rm {\bf Y}}_2 $ from one (or more) common variable $Y$, and ${\rm {\bf X}}_1 $ and ${\rm
{\bf X}}_2 $ from (common or non-common)  auxiliary variables with
known population totals $t_{X_1 } $ and $t_{X_2 } $, respectively. Let ${\rm {\bf w}} = \left( {{\rm {\bf {w_1^\text{T}}}} ,{\rm {\bf {w_2^\text{T}}}} }
\right)^\text{T} $ be the sampling weights, where $\textbf{w}_i=\left(w_{i,1},\ldots,w_{i,n_i}\right)$, $i=1,2$, and ${\rm {\bf a}}_\xi = \left( {{\rm {\bf
{a_{1,\xi}^\text{T}}}} ,{\rm {\bf {a_{2,\xi}^\text{T}}}} } \right)^\text{T} $ be the alignment weights, where $\xi$ stands for the alignment method. 

In the alignment method of \cite{merkouris2004}, the weights ${\rm {\bf a}}_{1,\xi} $ and ${\rm {\bf
a}}_{2,\xi} $ satisfy simultaneously the
constraints:
\[\widehat {Y_1^\xi} = {\rm {\bf Y}}_1^{\text{T}} {\rm {\bf a}}_{1,\xi} = {\rm {\bf Y}}_2^{\text{T}} {\rm {\bf a}}_{2,\xi} = \widehat {Y_2^\xi}
\]
and
\[\widehat {X_1^\xi} = {\rm {\bf X}}_1^{\text{T}} {\rm {\bf a}}_{1,\xi} = t_{X_1 }
, \quad \widehat {X_2^\xi} = {\rm {\bf X}}_2^{\text{T}}{\rm {\bf a}}_{2,\xi}=
t_{X_2 }.\]

The weights ${\rm {\bf a}}_{1,\xi} $ and ${\rm {\bf
a}}_{2,\xi} $  are obtained minimizing a generalized distance to the original weights ${\rm {\bf w}}_1 $ and ${\rm {\bf w}}_2 $,  that is minimizing 
\[\left( {{\rm {\bf a}}_\xi - {\rm {\bf w}}} \right)^{\text{T}} \boldsymbol{\Pi}
^{ - 1}\left( {{\rm {\bf a}}_\xi - {\rm {\bf w}}} \right),\]
where  $\boldsymbol{\Pi} =
\text{diag}\left( {\left( {{d_i } \mathord{\left/ {\vphantom {{d_i } {n_i
}}} \right. \kern-\nulldelimiterspace} {n_i }} \right)\boldsymbol{\Pi}_i} \right)$, $i=1,2$ is a block diagonal matrix, $d_i$ is the design effect associated with sample $S_i$, and $\Pi _i $ is a diagonal matrix with the $k$-th diagonal element to be equal to $w_{i,k} / q_{i,k}$ for some positive constants $q_{i,k}$. 

Thus, ${\rm {\bf a}}_\xi = \left( {{\rm {\bf{a}_{1,\xi}^{\text{T}}}} ,{\rm {\bf{a}_{2,\xi}^{\text{T}}}} } \right)^\text{T} $ are chosen so that not only auxiliary variables are calibrated to their known totals, but also, so that common target variables have aligned (identical) estimates across samples. Note that the alignment weights,  together with the data from one sample only, contain the information available in all samples about the common variables. The advantage of such a
procedure is that it allows improving even upon the estimates of variables which are not common to the samples. Their estimates, being based on the alignment weights, exploit the correlation of these variables with the common variables.

The alignment weights ${\rm {\bf a}}_\xi = \left( {{\rm {\bf{a}_{1,\xi}^{\text{T}}}} ,{\rm {\bf{a}_{2,\xi}^{\text{T}}}} } \right)^\text{T} $ may be calculated as \[{\rm {\bf a}}_\xi = {\rm {\bf w}}
+  \boldsymbol{\Pi} {\cal \boldsymbol{N}} \left( {{\cal \boldsymbol{N}}^\textbf{T} \boldsymbol{\Pi} {\cal \boldsymbol{N}} } \right)^{ -
1}\left( {\textbf{t}_X - {\cal \boldsymbol{N}}^\text{T} {\rm {\bf w}}} \right),\]
 where \[
 {\cal N}= \left(
{{\begin{array}{*{20}c}
 {{\rm {\bf X}}_1 } & 0 & {{\rm {\bf Y}}_1 } \\
 0 & {{\rm {\bf X}}_2 } & { - {\rm {\bf Y}}_2 } \\
\end{array} }} \right)\]  and \[ t_X = \left( {{\begin{array}{*{20}c}
 {t_{X_1 } } \hfill & {t_{X_2 } } \hfill & 0 \hfill \\
\end{array} }} \right)^\text{T} . \] 

More specifically, we get that
\[ {\rm {\bf a}}_{1,\xi} = {\rm {\bf
c}}_1 + \left( {1 - \phi } \right)\Xi _1 {\rm {\bf Y}}_1 \left[
{\left( {1 - \phi } \right){\rm {\bf Y}}_1^\text{T} \Xi _1 {\rm {\bf
Y}}_1 + \phi {\rm {\bf Y}}_2^\text{T} \Xi _2 {\rm {\bf Y}}_2 }
\right]^{ - 1}\left( {\widehat {Y_2^R} - \widehat {Y_1^R} } \right) \]
and
\[ {\rm {\bf a}}_{2,\xi} = {\rm {\bf
c}}_2 - \phi\Xi _2 {\rm {\bf Y}}_2 \left[
{\left( {1 - \phi } \right){\rm {\bf Y}}_1^\text{T} \Xi _1 {\rm {\bf
Y}}_1 + \phi {\rm {\bf Y}}_2^\text{T} \Xi _2 {\rm {\bf Y}}_2 }
\right]^{ - 1}\left( {\widehat {Y_2^R} - \widehat {Y_1^R} } \right), \]

where 
\[ \textbf{c}_i = \textbf{w}_i + \boldsymbol{\Pi}_i {\bf X}_i \left(
{\bf X}_i^\text{T} \boldsymbol{\Pi}_i {\bf
X}_i\right)^{ - 1}\left( t_{X_i } - {\bf X}_i\textbf{w}_i^\text{T} \right),
\]
\[ \phi = {\left( {{n_1 }
\mathord{\left/ {\vphantom {{n_1 } {d_1 }}} \right.
\kern-\nulldelimiterspace} {d_1 }} \right)} \mathord{\left/
{\vphantom {{\left( {{n_1 } \mathord{\left/ {\vphantom {{n_1 }
{d_1 }}} \right. \kern-\nulldelimiterspace} {d_1 }} \right)}
{\left[ {\left( {{n_1 } \mathord{\left/ {\vphantom {{n_1 } {d_1
}}} \right. \kern-\nulldelimiterspace} {d_1 }} \right) + \left(
{{n_2 } \mathord{\left/ {\vphantom {{n_2 } {d_2 }}} \right.
\kern-\nulldelimiterspace} {d_2 }} \right)} \right]}}} \right.
\kern-\nulldelimiterspace} {\left[ {\left( {{n_1 } \mathord{\left/
{\vphantom {{n_1 } {d_1 }}} \right. \kern-\nulldelimiterspace}
{d_1 }} \right) + \left( {{n_2 } \mathord{\left/ {\vphantom {{n_2
} {d_2 }}} \right. \kern-\nulldelimiterspace} {d_2 }} \right)}
\right]}, \]
\[
\boldsymbol{\Xi} _i = \boldsymbol{\Pi} _i \left[ {{\rm {\bf I}} - {\rm {\bf X}}_i \left(
{{\rm {\bf {X}^\text{T}}}_i \boldsymbol{\Pi} _i {\rm {\bf X}}_i } \right)^{ - 1}{\rm
{\bf {X}^\text{T}}}_i \boldsymbol{\Pi} _i } \right] \] 
and 
\begin{equation}\label{y_hat}
\widehat {Y_i^R} = \widehat {Y_i} + \widehat {\beta _i} \left( {t_{X_i } - {\bf X}_i\textbf{w}_i^\text{T} } \right).
\end{equation}
\noindent
$\widehat {Y_1}$ and $\widehat {Y_2}$ are the Horvitz-Thomspon estimators of the totals and they are given by $\bf{Y}_1\textbf{w}_1^\text{T}$ and $\bf{Y}_2\textbf{w}_2^\text{T}$, respectively. $\widehat {Y_1^R}$ and $\widehat {Y_2^R}$ are the regression estimates from $S_1$ and $S_2$ calibrated on $X_1$ and $X_2$, respectively. $\widehat {\beta _1}$ and $\widehat {\beta _2}$ are the corresponding estimated coefficients from the generalized regression estimators for which we have that $\widehat {\beta _i} = {\rm {\bf {Y}_i^\text{T}}} \boldsymbol{\Pi}
_i {\rm {\bf X_i^\text{T}}} \left( {{\rm {\bf {X}_i^\text{T}}} \boldsymbol{\Pi}_i {\rm {\bf
X}}_i } \right)^{ - 1}$. Expression \eqref{y_hat} indicates that $\widehat {Y_i^R} $ implicitly  uses the correlation of the variable of interest $Y$ with the auxiliary variable $X_i$, together with the fact that $t_{X_i} $ is
known, in order to correct $\widehat {Y_i} $ by an appropriate amount.

Therefore, we get the estimator
\[ \widehat {Y_1^{\xi}} = {\rm {\bf Y}}_1^\text{T} {\rm {\bf a}}_{1,\xi} = \widehat
{Y_1^R} + \left( {1 - \widehat {\gamma ^\xi}} \right)\left( {\widehat
{Y_2^R} - \widehat {Y_1^R} } \right), \] which is identical to the estimator \[ \widehat
{Y_2^{\xi}} = {\rm {\bf Y}_2^\text{T}} {\rm {\bf a}}_{2,\xi} = \widehat {Y_2^R} -
\widehat {\gamma ^\xi}\left( {\widehat {Y_2^R} - \widehat {Y_1^R} } \right),\]
that is \[ \widehat {Y_1^{\xi}} = \widehat {Y_2^{\xi}} = \widehat {\gamma
^\xi}\widehat {Y_1^R} + \left( {1 - \widehat {\gamma ^\xi}} \right)\widehat
{Y_2^R}, \] 
where 
 \begin{equation}\label{for_align}
 \widehat {\gamma ^\xi} = \phi {\rm {\bf
Y_2^\text{T}}} \boldsymbol{\Xi}_2 {\rm {\bf Y_2}} \left[ {\left( {1 - \phi }
\right){\rm {\bf Y_1^\text{T}}} \boldsymbol{\Xi}_1 {\rm {\bf Y_1}} + \varphi {\rm
{\bf Y_2^\text{T}}} \boldsymbol{\Xi}_2 {\rm {\bf Y_2}} } \right]^{ - 1}.     
 \end{equation}

Hence expression \eqref{for_align} relates
the alignment estimates to those from standard calibration.

The estimates of variances for large samples are given by
\[\widehat {Var}\left( {\widehat {Y_2^{\xi}} } \right) = \boldsymbol{\Psi}_{\xi}^\text{T}
\boldsymbol{\Pi}_0\boldsymbol{\Psi}_\xi,\] 
where 
\[\boldsymbol{\Psi}_\xi = \left[ {\widehat {\gamma ^\xi}\left(
{{\rm {\bf {Y}^\text{T}_1}} - \widehat {\beta _1} {\rm {\bf {X}_1^\text{T}}} }
\right),\left( {1 - \widehat {\gamma ^\xi}} \right)\left( {{\rm {\bf
{Y}_2^\text{T}}} - \widehat {\beta _2} {\rm {\bf {X}_2^\text{T}}} } \right)} \right]^\text{T}
\]
and $\boldsymbol{\Pi}_0$ is a semi positive-definite matrix whose $kl$-th entry is equal to $\left(\pi_{k \ell}-\pi_k\pi_{\ell}\right)/\pi_k\pi_{\ell}$ ($\pi_{kk}\equiv\pi_{k}$).

Finally, there is no a clear rule to define what constitutes a ``large
sample". However, the size of social surveys is typically
considered large enough, unless a very large number of auxiliary
variables are used in generalized regression (GREG) or unless estimates are produced for
very small domains.This provides all the necessary tools to obtain the estimates.

\section{Complex statistics}\label{complex}
In the present paper we focus on complex statistics from social surveys related to poverty estimation. 
In this section, we provide some details about the statistics themselves and  the linearized variables for the complex statistics addressed in this paper.

Let $y_k$ be a given characteristic associated to each unit $u_k$.The total in the finite population $U$ is defined by $Y = \sum_{k \in U} y_k$. The estimators of the population size and the total are respectively defined by $\widehat N = \sum_{k \in S} w_k$ and $\widehat Y = \sum_{k \in S} w_k y_k$. Note that the weights $w_k$  can be of any kind, for example based on the sampling scheme or via calibration.  

\subsection{Median}
The median is considered as a more appropriate measure of location than the mean when dealing with variables of interest, such as income and expenditure, that have highly skewed distributions. This is because the median is less sensitive to outliers compared to the mean. For this reason, the median is also used by most household wealth surveys,
such as the Household Finance and Consumption Survey (HFCS) carried out by the European Central Bank among the Eurozone countries.
 
When the sample design is complex (not simple random sampling), as expected, the estimation of the cumulative distribution function and consequently, quantiles, becomes more complicated. The reason is that the simple proportion of values lower or equal than the desired value can have bad behavior and hence improvement is needed.

Let the data $y_i$ be sorted into ascending order. Also, let $w_i$
be the weight associated to the $i$-th value, where $i \in S$ and $S$ is the sample. Note that the weights can be of any kind,
for example based on the sampling scheme or via calibration. Then
the estimated $a$-quantile is
\[
\widehat{Q_a} = \left\{ \begin{array}{lc} \frac{1}{2}(y_j+y_{j+1}) &
\mbox{if } \sum\limits_{i=1}^j w_i = a \widehat N \\
y_{j+1} & \mbox{if } \sum\limits_{i=1}^j w_i < a \widehat N  <
\sum\limits_{i=1}^{j+1} w_i
\end{array}. \right.
\]
This is the typical quantile
estimator but using the weights for each observation.

For the median, the linearized variable $z_k$ is estimated from $S$ by its plug-in estimator 
\begin{equation}
\widehat{z_k^m} = - \frac{1}{ \widehat{f(\widehat{Q_{0.5}})} } \frac{1}{\widehat N} \left[
\textbf{1}(y_k \le \widehat{Q_{0.5}}) -0.5\right], \label{linmed}
\end{equation}
where $\textbf{1}(\cdot)$ is the usual indicator function and $\widehat{ f(\widehat{Q_{0.5}})}$ is the estimated density at $\widehat{Q_{0.5}}$. This density estimator can be a kernel based one \citep{graf2014},
\[
\widehat{f(x)} = \frac{1}{h \widehat N} \sum\limits_{k \in S} w_k K\left(
\frac{x-y_k}{h}\right),
\]
where $h$ is the bandwidth that determines the amount of smoothing of
the data, and it is obtained from the data as $\widehat h = \widehat{\sigma}\widehat
{N}^{-0.2}$, where $\widehat \sigma$ is an estimator of the variance of
the variable $y$. A natural choice for kernel is the standard
normal distribution, i.e. $K(\cdot)$ is the pdf of the standard
normal distribution. More details can be found in \cite{graf2014}.

It is interesting to point out that by
construction the linearized variable will take only two values,
whether they are above or below the median.

So, in order to obtain an estimate for the variance of the median, one needs to
calculate the linearized variable given in
(\ref{linmed}) and then to use (\ref{variance}) to
get the variance estimate.

\subsection{Quintile share ratio S80/S20}
The quintile share ratio S80/S20 (QSR) has become a very popular measure of social inequality and cohesion \citep{atkinson2004indicators}.  The QSR is generally defined as the ratio of the total income earned by the richest 20\% of the population relative
to that earned by the poorest 20\%. Thus,it can be expressed using quantile shares, where a quantile share is the share of total income earned by all of the units up to a given quantile.

Consider a strictly increasing cumulative distribution function $F(y)$, while $f(y)$ is its density function.  A quantile share is the share of total income earned by all the income earners up to quantile of order $a$. Namely we have that
\[
S(a) = \frac{\int_0^\xi u dF(u)}{\int_0^\infty u dF(u)}.
\]
This quantity is also known as the Lorenz curve (or function) and is crucial in inequality calculation. Recall that the Gini coefficient is also defined using this concept.

The
finite population quantile share is
\[
S(a) = \frac{Y_a}{Y},
\]
where $Y_a = \sum_{k \in U} y_k \textbf{1}(y_k \le Q_a)$ \citep{langel2011}. 

As we have already mentioned, while the definition of a quantile in a
continuous distribution is clear and unique, it is not so in the case of finite population context, where $F(\cdot)$, the cumulative distribution function of the income is usually a step function.

Based on all the aforementioned, the QSR is defined as
\[
QSR= \frac{1-S(0.8)}{S(0.2)},
\]
and its estimate for a finite population is
\[
{\widehat {QSR}} = \frac{\widehat Y - \widehat{Y_{0.8}}}{\widehat {Y_{0.2}}},
\]
where $\widehat {Y_a} = \sum_{k \in S} w_k y_k \textbf{1}(y_k \le \widehat {Q_a})$.

For the QSR, the linearized variable $z_k$ is estimated from $S$ by its plug-in estimator 
\begin{equation} 
\begin{split}
\widehat{z_k^{QSR}} & = \frac{y_k - 0.8\widehat{Q_{0.8}}+(\widehat{Q_{0.8}}-y_k)\textbf{1}(y_k \le \widehat{Q_{0.8}})}{\widehat{Y_{0.2}}}\\ & - 
\frac{(\widehat{Y} - \widehat{Y_{0.8}})\left[0.2\widehat{Q_{0.2}}-(\widehat{Q_{0.2}}-y_k)\textbf{1}(y_k \le \widehat{Q_{0.2}})\right]}{{\widehat {Y_{0.2}}}^2}.
\label{linQSR} 
\end{split}
\end{equation}

So, in order to obtain an estimate for the variance of the QSR, one needs to
calculate the linearized variable given in
(\ref{linQSR}) and then to use (\ref{variance}) to
get the variance estimate.

\subsection{Gini coefficient} 
The Gini coefficient has been one of the most widely used measures of inequality. A significant amount of work has already been conducted on this measure both from theoretical and empirical part.

If a population $U$ is finite, the Gini coefficient can be calculated as
\[
G= \frac{2 \sum_{k \in U} k y_k}{N \sum_{k \in U} y_k} -
\frac{N+1}{N},
\]
where $y_k$ are sorted. For a sample, the Gini
coefficient can be estimated as \begin{eqnarray*} \widehat G & =&
\frac{2}{\widehat N \widehat Y} \sum_{k \in S} w_k \widehat {N_k} y_k -
\left( \frac{1}{\widehat N \widehat Y} \sum_{k \in S} w_k^2 y_k \right) \\
 &=& \frac{\sum_{k
\in S} \sum_{\ell \in S} w_k w_\ell |y_k - y_\ell|}{2 \widehat N \widehat
Y},
\end{eqnarray*}
where $\widehat{N_k} = \sum_{\ell \in S}  w_\ell \textbf{1}(y_\ell \le y_k)$. In case of equal weights the expression of the estimate of the Gini coefficient can be simplified.

For the Gini coefficient, the linearized variable $z_k$ is estimated from $S$ by its plug-in estimator 
\begin{equation} \widehat{z_k^G} = \frac{1}{\widehat N \widehat Y} \left[ 2 \widehat{ N_k}
(y_k - \widehat{\bar Y})+ \widehat Y - \widehat N y_k - \widehat G (\widehat Y + y_k
\widehat N)) \right], \label{linGin} \end{equation} where $\widehat {\bar
Y} = \widehat
Y/\widehat{N_k}$.

So, in order to obtain an estimate for the variance of the Gini coefficient, one needs to
calculate the linearized variable given in
(\ref{linGin}) and then to use (\ref{variance}) to
get the variance estimate.

\subsection{Relative median at-risk-of-poverty gap}

The relative median at-risk-of-poverty gap is calculated as the
difference between
 the median equivalized disposable income of people below the at-risk-of-poverty threshold and the at-risk-of-poverty threshold,
 expressed as a percentage of the at-risk-of-poverty threshold
 (cut-off point: 60\% of national median equivalized disposable income). The equivalized disposable income is the total income of a household, after tax and other deductions, that is available for spending or saving, adjusted by the number of equalised adults of the household.

The at-risk-of-poverty threshold (ARPT) is defined as 60\% of the
median income, i.e.
\[
\widehat {ARPT} = 0.6 \widehat{Q_{0.5}}.
\]
The at-risk-of-poverty rate (ARPR) is the proportion  of the
population with an income below the ARPT. It takes values (being a proportion) at $[0,1]$. The ARPR is scale-independent, like the Gini coefficient and the  QSR.  The definition of the estimated ARPR for a sample \citep{fabrizi2020} is
\[
\widehat{ARPR} = \frac{\sum\limits_{k \in S} w_k \textbf{1}(y_k \le
\widehat{ARPT}) }{\widehat N}.
\]

The estimator of the median income of individuals below the ARPT is denoted as
$\widehat{m_p}$ and it is estimated in the same way as any other
quantile but using only the data points that are below the ARPT.

Now we can define the relative median at-risk-of-poverty gap
(RMPG) as
\[
\widehat {RMPG} = \frac{ \widehat {ARPT}- \widehat{m_p}}{ \widehat
{ARPT}}.
\]

It is clear that the estimator depends on the ARPT and the ARPR as well as involves the estimation of another  median.

For the RMPG \citep{osier2009,graf2014}, the linearized variable $z_k$ is estimated from $S$ by its plug-in estimator  
\begin{equation} \widehat{z_k^{RMPG}} = \frac{\widehat{m_p}
\widehat{z_k^{ARPT}} - \widehat{ARPT} \widehat{z_k^{\widehat{m_p}}} }{\widehat{ARPT}^2},
\label{linRMPG}
\end{equation}
where $\widehat{z_k^{ARPT}}$ and $\widehat{z_k^{\widehat{m_p}}}$ are the plug-in estimators from $S$ for the linearized
variable $z_k$ for the ARPT and the $\widehat{m_p}$, respectively. 

For the ARPT, the linearized variable $z_k$ is estimated from $S$ by its plug-in estimator 
\[
\widehat{z_k^{ARPT}} = - \frac{0.6}{ \widehat{f(\widehat{Q_{0.5}})} } \frac{1}{\widehat N}
\left[ \textbf{1}(y_k \le \widehat{Q_{0.5}}) -0.5 \right],
\]
and given that for the ARPR the linearized variable $z_k$ is estimated from $S$ by its plug-in estimator 
\[
\widehat{z_k^{ARPR}} = \frac{1}{\widehat N} \left(\textbf{1}(y_k \le
\widehat{ARPT})-\widehat{ARPR} \right) + \widehat{f( \widehat{ARPT})}
\widehat{z_k^{ARPT}},
\]
for $\widehat{m_p}$ the linearized variable $z_k$ is estimated from $S$ by its plug-in estimator 
\[
\widehat{z_k^{\widehat{m_p}}} = \frac{1}{\widehat{f(\widehat{m_p})}} \frac{\widehat{z_k^{ARPR}}}{2}
- \frac{1}{\widehat N} \left[ \textbf{1}(y_k \le \widehat{m_p})- F(\widehat{m_p}) \right].
\]

So, in order to obtain an estimate for the variance of the RMPG, one needs to
calculate the linearized variable given in
(\ref{linRMPG}) and then to use (\ref{variance}) to get the variance estimate.

\section{Alignment weights derived from linearized variables}\label{our_approach}
In this section, the proposed approach of estimating complex statistics pooling information from different surveys through the alignment method of \cite{merkouris2004} is described. Even though our approach is implemented for the complex statistics provided in Section \ref{complex}, its application to other complex statistics is straightforward, such as utilizing the linearized variable of the Bonferroni inequality index \citep{dong2021}.

The main obstacle for applying the alignment method is the non-linear form of the complex statistics, Therefore, we propose the usage of the linearized variable for every complex statistic. The key property given in \cite{deville1999} is that a complex statistic can be written as
\begin{equation}\label{eq_dev}
 \widehat \theta - \theta \approx \sum\limits_{k \in S} z_k w_k -
\sum\limits_{k \in U} z_k.
\end{equation}
This representation via a total is the key ingredient for variance  calculations.

We consider two samples $S_1$ and $S_2$ as well as their complex statistic estimates derived from them, say $\widehat{\theta_1}$ and $\widehat{\theta_2}$.  Based on \eqref{eq_dev}, the alignment method aims at equating the two statistics, that is setting $\widehat{\theta_1} = \widehat
\theta_2$. Since the quantities $\sum\limits_{k \in U} z_k$ and
$\theta$ are population values, it is clear that we get 
\[
\sum\limits_{k \in S_1} z_k w_k  \approx \sum\limits_{k \in S_2}
z_k w_k,
\]
implying that we have to align the linearized variable related to the complex statistic. Therefore, the approach of \cite{merkouris2004} is applicable and can be implemented, given that we are now dealing with totals. Consequently, we can derive alignment weights for each sample and proceed similarly to the alignment method for totals. We recall that the variances estimates of $\widehat{\theta_1}$ and $\widehat{\theta_2}$ can be derived using \eqref{variance}.

Once the alignment weights ${\rm {\bf a}}_\xi = \left( {{\rm {\bf
{a_{1,\xi}^\text{T}}}} ,{\rm {\bf {a_{2,\xi}^\text{T}}}} } \right)^\text{T} $ have been obtained, then the variance estimates for the new alignment estimators, say $\widehat{\theta_{1,\xi}}$ and $\widehat{\theta_{2,\xi}}$, can be obtained using the linearized variable derived based on the alignment weights, that is using \eqref{variance} we get

\begin{equation}
{\widehat{Var}}(\widehat{\theta_{1,\xi}}) = \sum\limits_{k \in S_1} \sum\limits_{\ell
\in S_1} \tilde \Delta_{k \ell}  \frac{\widehat{z_{k,\xi}} \widehat{z_{\ell,\xi}}}{ \pi_k \pi_\ell}
\label{variance_theta1}
\end{equation}
and
\begin{equation}
{\widehat{Var}}(\widehat{\theta_{2,\xi}}) = \sum\limits_{k \in S_2} \sum\limits_{\ell
\in S_2} \tilde \Delta_{k \ell}  \frac{\widehat{z_{k,\xi}} \widehat{z_{\ell,\xi}}}{ \pi_k \pi_\ell},
\label{variance_theta2}
\end{equation}
where $\widehat{z_{k,\xi}}$ is the plug-in estimator for the linearized variable $z_k$ using the alignment weights.

Moreover, let $\widehat{\theta_c}$ be the estimator of $\theta$ which is obtained by combining the estimators $\widehat{\theta_{1,\xi}}$ and $\widehat{\theta_{2,\xi}}$. Thus, we have that 
\begin{equation}
\widehat{\theta_c}=\delta\widehat{\theta_{1,\xi}}+(1-\delta)\widehat{\theta_{2,\xi}},
\label{theta_c}
\end{equation}
where $\delta\in[0,1]$. Then, the estimated variance of $\widehat{\theta_c}$ is
\begin{equation*}
{\widehat{Var}}(\widehat{\theta_c})=\delta^2{\widehat{Var}}(\widehat{\theta_{1,\xi}})+(1-\delta)^2{\widehat{Var}}(\widehat{\theta_{2,\xi}}).
\label{variance_theta_c}
\end{equation*}

If in \eqref{theta_c} we set $\delta=0.5$, then $\widehat{\theta_c}$ is a simple average of $\widehat{\theta_{1,\xi}}$ and $\widehat{\theta_{2,\xi}}$. An optimal value for $\delta$ can be derived using a weighted average based on $\widehat{Var}(\widehat{\theta_{1,\xi}})$ and $\widehat{Var}(\widehat{\theta_{2,\xi}})$, that is 
\begin{equation}
\delta_{opt}=\dfrac{\widehat{Var}(\widehat{\theta_{2,\xi}})}{\widehat{Var}(\widehat{\theta_{1,\xi}})+\widehat{Var}(\widehat{\theta_{2,\xi}})}.
\label{optimal_delta}
\end{equation}

In Algorithm 1 we outline the steps of our proposed method for estimating complex statistics by combining different surveys.

\begin{algorithm}
	\caption{}
	\begin{algorithmic}
		\State \textbf{Step 1:} Obtain the plug-in estimator $\widehat{z_k}$ for the linearized variable $z_k$ associated with the complex statistic under consideration using the weights $w_k$ derived according to the sampling scheme.
		\State \textbf{Step 2:} Apply the alignment method using the plug-in estimator $\widehat{z_k}$ from Step 1 to obtain the alignment weights ${\rm {\bf a}}_\xi = \left( {{\rm {\bf
{a_{1,\xi}^\text{T}}}} ,{\rm {\bf {a_{2,\xi}^\text{T}}}} } \right)^\text{T} $.
		\State \textbf{Step 3:} Compute the estimators $\widehat{\theta_{1,\xi}}$ and $\widehat{\theta_{2,\xi}}$ as well as their variances from \eqref{variance_theta1} and \eqref{variance_theta2} using the alignment weights  ${\rm {\bf a}}_\xi = \left( {{\rm {\bf
{a_{1,\xi}^\text{T}}}} ,{\rm {\bf {a_{2,\xi}^\text{T}}}} } \right)^\text{T} $ obtained in Step 2.
		\State \textbf{Step 4:} Obtain the weighted estimator $\widehat{\theta_c}$ for the considered complex statistic according to the estimators  $\widehat{\theta_{1,\xi}}$ and $\widehat{\theta_{2,\xi}}$ 
 and their variances computed in Step 3.
	\end{algorithmic}
\end{algorithm}

\section{Application to the \texttt{eusilc} dataset}\label{results}
The basis for the simulations are the synthetic population data generated from Austrian European
Union Statistics on Income and Living Conditions (EU-SILC) data from 2006, the latter of which were provided by Statistics Austria. The population data are thereby simulated with the methodology described in \cite{alfons2011} and implemented in the R package \texttt{simPopulation} \citep{alfons_package}. 

The package \texttt{laeken} \citep{alfons2013} provides the synthetic example dataset named \texttt{eusilc} consisting of 14,827 observations with 28 variables from $6,000$
households. Not all the variables are useful but some of them will be used
later on for calibration-alignment purposes. 

In both Subsections \ref{one-sample} and \ref{two-sample} the simulation is repeated $1,000$ times. We report results based on these 1,000 replications. 

\subsection{One-sample results}\label{one-sample}
The results refer to two-stage sampling, by sampling households $1,000$ out of $6,000$ available ones, and then taking into consideration all individuals from the household. This sampling scheme allows for an easier calculation of the second-order probabilities that we need to use for illustrating the potential of the developed methodology. We need to mention that we consider as population the entire \texttt{eusilc} dataset.

For such a sampling scheme we can derive the second-order inclusion probabilities as follows. Let $H$ be the number of the available households. For the $j$-th household we have $n_H$ individuals, and so $\sum n_H = N$ is the total population size. We sample randomly $n_h$ households. Then, the first-order inclusion probabilities are $n_h/H$ since we have simple random sampling
on this stage. For the second stage we need the probability of joint inclusion of any two individuals. This is $n_h/H$ for individuals on the same household and $\frac{n_h}{H}\frac{n_h-1}{H-1}$ for those not in the same household. This provided relatively easy the second-order inclusion probabilities needed for the variance calculation given in (\ref{variance}).

Table \ref{tab1} shows the estimated quantities and their standard error based on a random sample of $1,000 $ households.

\begin{table}[ht]
\begin{center}
\begin{tabular}{crr}
\hline
Indicator & Estimate & Standard error \\
\hline Median & 18107.49  & 326.9414 \\
QSR & 3.8734 & 0.1457 \\
Gini coefficient & 26.2626 & 0.6785\\
RMPG (\%)& 18.8592 & 2.3052 \\\hline
\end{tabular}
\caption{\label{tab1} Estimated quantities and the standard error of four indicators, based on a sample size of $1,000$ households and a two-stage sampling design.
}
\end{center}
\end{table}

\begin{figure}[ht]
\begin{center}
\includegraphics[scale=0.57]{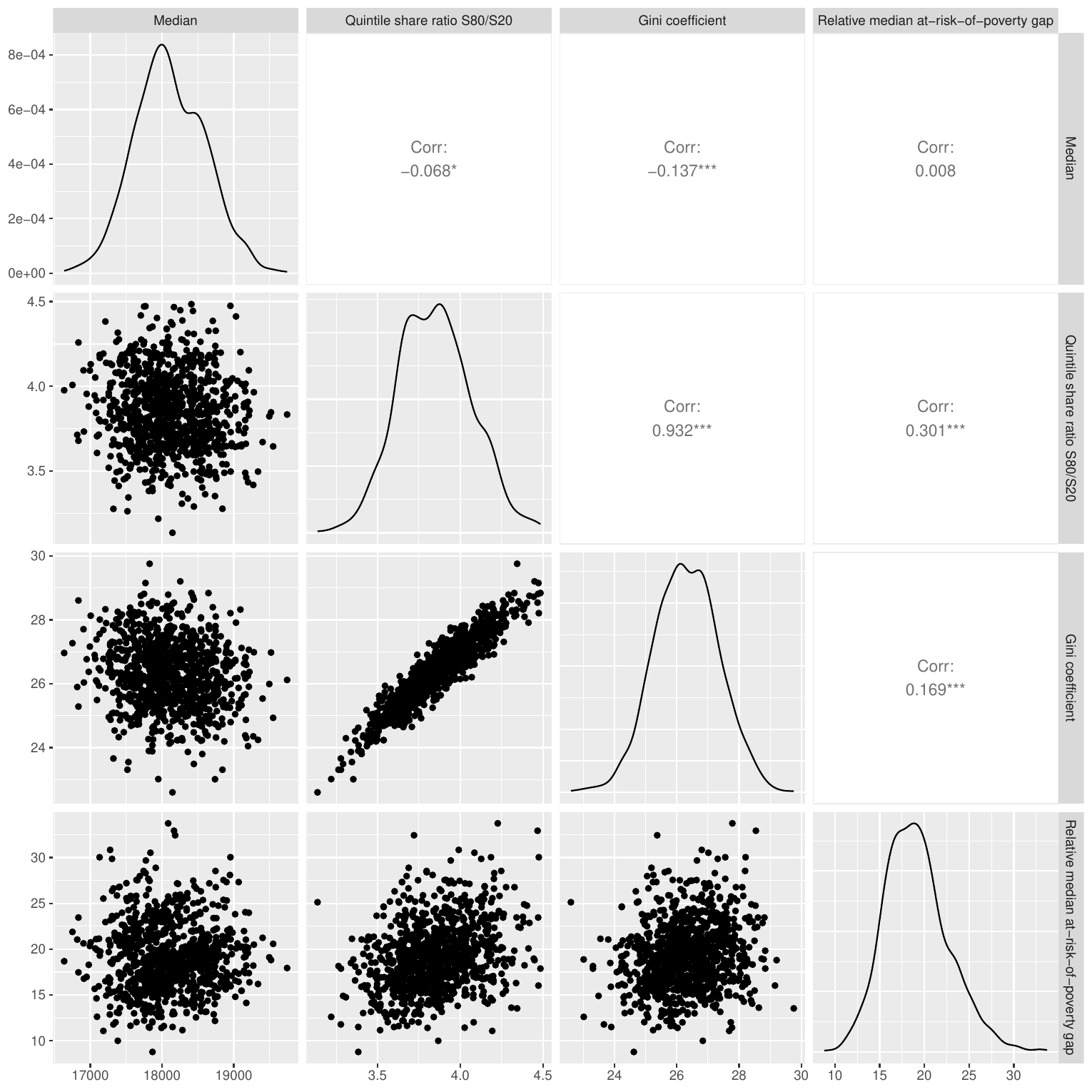}
\caption{\label{fig1} The relationships between the estimators of the indicators given in Table \ref{tab1}, based on a sample size of $1,000$ households and a two-stage sampling design.}
\end{center}
\end{figure}

Figure \ref{fig1} shows the relationships between the estimators based on sample size of
$1,000$ households, across the $1,000$ replications. It is interesting that QSR is strongly correlated to Gini coefficient and this is the case for many replications of such an experiment.

\begin{figure}[ht]
\begin{center}
\includegraphics[width=1\textwidth]{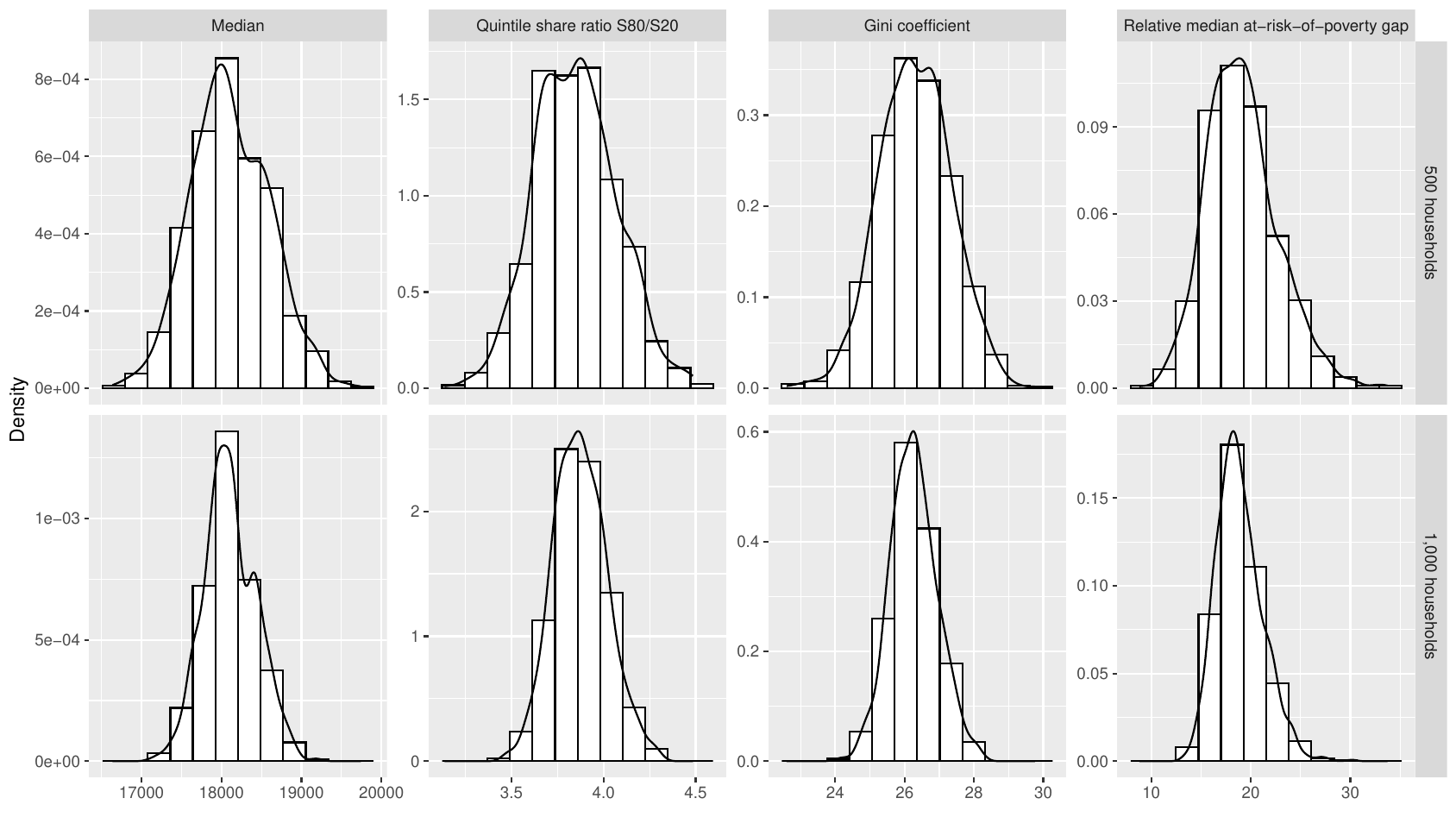}
\caption{\label{fig2} Histograms and density plots for the four indicators given in Table \ref{tab1}, based on a sample size of $500$ and $1,000$ households, and a two-stage sampling design.}
\end{center}
\end{figure}

Based on a two-stage sampling design, Figure \ref{fig2} shows the histograms and the density plots for the four indicators given in Table \ref{tab1} for a sample size of $500$ and $1,000$ households. All estimators show some skewness, especially the RMPG. The histogram in Figure \ref{fig2} aims at revealing the change on the variability from $500$ to $1,000$ households. Considering the variability from the histograms and the density plots as a Monte Carlo based estimate of the variance, we see that it performs relatively well.

\subsection{Two-sample results}\label{two-sample}
Now consider that we have two samples and we want to pool information from them. 
In order to apply the alignment method, we examine three scenarios about the sizes of the two samples, among which there is no overlapping. Let $n_{h_i}$ be the size of the $i$-th sample, i.e. the number of the randomly sampled households. Note that we take into consideration all individuals within the household. In the first scenario, we have $n_{h_1}=500$ and $n_{h_2}=1,000$, in the second scenario we have $n_{h_1}=n_{h_2}=1,000$, and in the third scenario we have $n_{h_1}=3,000$ and $n_{h_2}=1,000$. The results for the second and the third scenarios are provided in the supplementary material, since they are similar to these of the first scenario.

From the dataset we assume that the variable \texttt{py010n} (employee cash or near cash income) is an auxiliary variable for which the total is known. We derive the ``true" values from the $6,000$ households, since we consider as population the entire dataset. The selected variable is correlated ($r=0.36$) with the income. Of course, this correlation is perhaps not a big one, but this is the larger correlation on the available dataset. Thus, we emphasize that we select this variable only for illustration purposes in order to apply our proposed methodology.

We apply the alignment method of complex statistics described in Section \ref{our_approach}  for the median, the QSR, the Gini coefficient and the RMPG taking into account the sampling
scheme. For the calculation of the estimators of the aformentioned complex statistics, we have three scenarios. In the first scenario, we use the weights that are based on the sampling scheme. In the second scenario, we use alignment weights by aligning for each statistic separately. In the third scenario, we use the alignment weights from the median. Moreover,  we combine the estimators of the two samples using two different methods. In the first method we use a simple average of the two samples, that is in \eqref{theta_c} we set $\delta=0.5$. In the second method, which is an optimal one, we use a weighted average of the two samples based on their variances, that is we use $\delta_{opt}$ given in \eqref{optimal_delta}.

\begin{figure}[!ht]
\begin{center}
\includegraphics[scale=0.65]{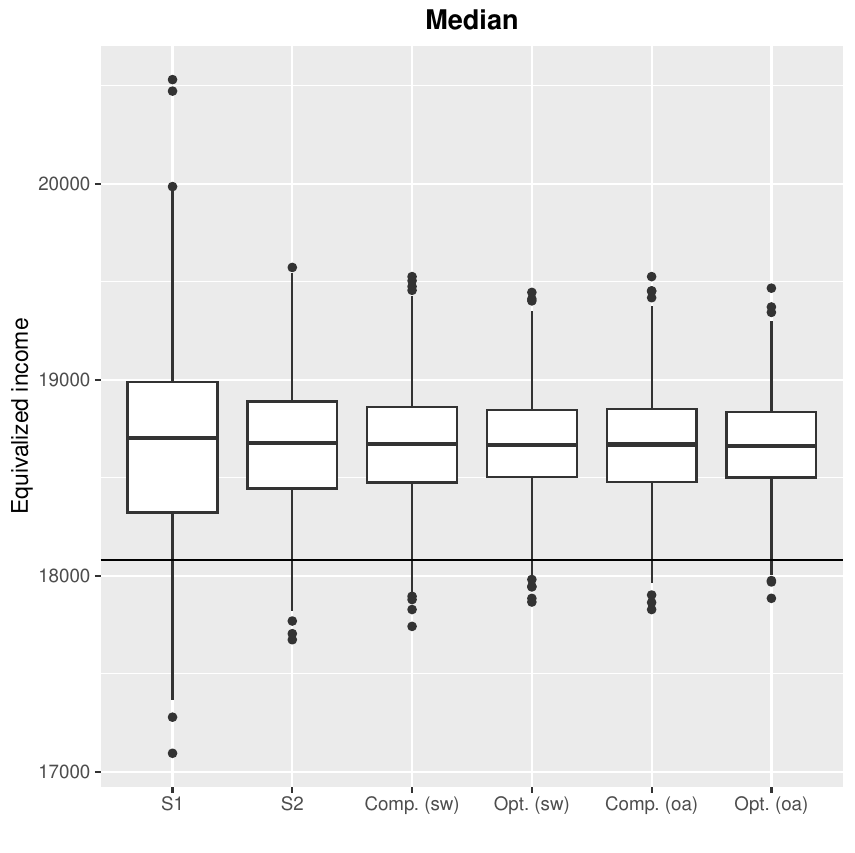}
\caption{\label{median_500_1000} The median of the equivalized income when $n_{h_1}=500$ and $n_{h_2}=1,000$. S1 and S2 stand for the first and the second sample, respectively. The weights are based on the sampling scheme (sw) and aligning the median (oa). The estimators of the two samples are combined using a simple average (Comp.) and in an optimal way (Opt.). The horizontal line is the ``true" value on the population.}
\end{center}
\end{figure}

\begin{figure}[!ht]
\begin{center}
\includegraphics[scale=0.65]{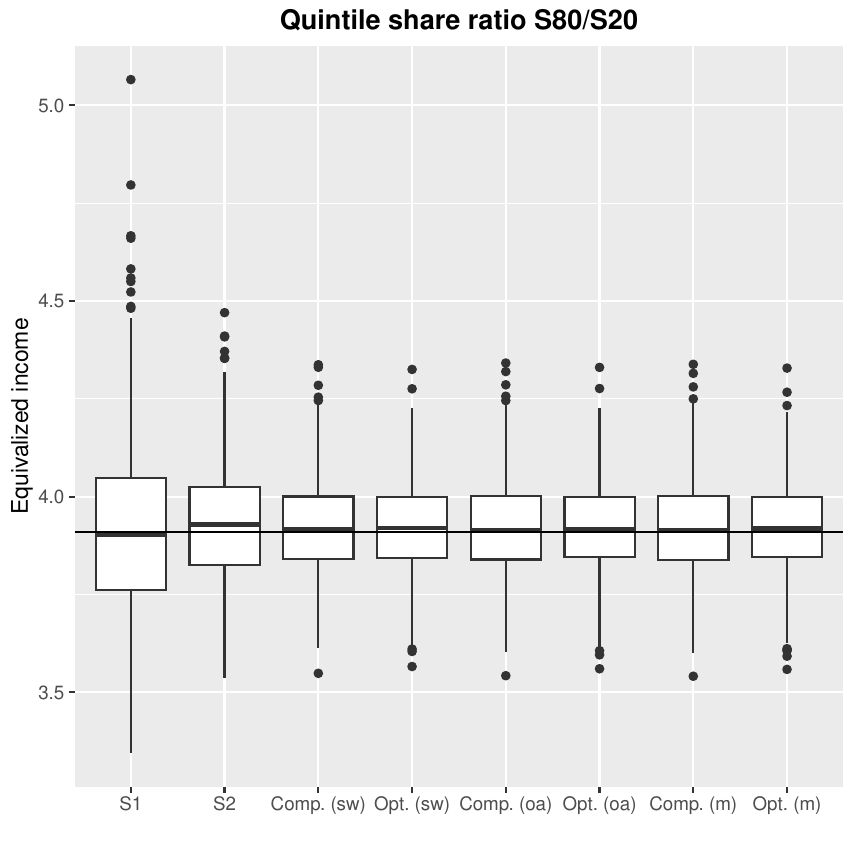}
\caption{\label{qsr_500_1000} The QSR of the equivalized income when $n_{h_1}=500$ and $n_{h_2}=1,000$. S1 and S2 stand for the first and the second sample, respectively. The weights are based on the sampling scheme (sw), aligning the QSR (oa), and the alignment of the median (m). The estimators of the two samples are combined using a simple average (Comp.) and in an optimal way (Opt.). The horizontal line is the ``true" value on the population.}
\end{center}
\end{figure}

\begin{figure}[!ht]
\begin{center}
\includegraphics[scale=0.65]{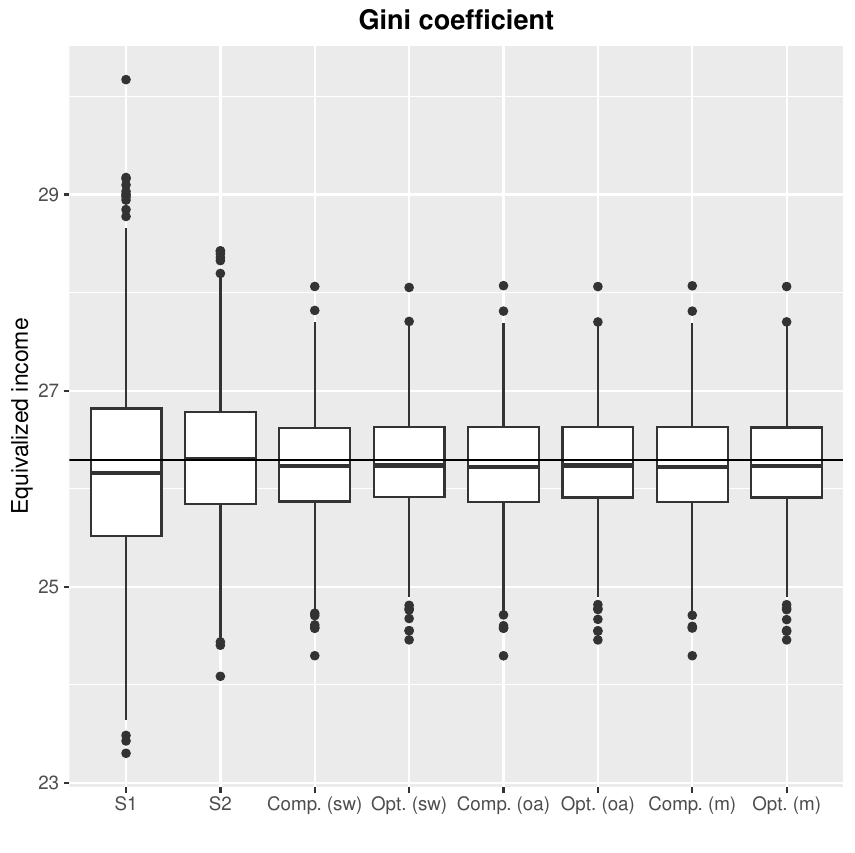}
\caption{\label{gini_500_1000} The Gini coefficient of the equivalized income when $n_{h_1}=500$ and $n_{h_2}=1,000$. S1 and S2 stand for the first and the second sample, respectively. The weights are based on the sampling scheme (sw), aligning the Gini coefficient (oa), and the alignment of the median (m). The estimators of the two samples are combined using a simple average (Comp.) and in an optimal way (Opt.). The horizontal line is the ``true" value on the population.}
\end{center}
\end{figure}

\begin{figure}[!ht]
\begin{center}
\includegraphics[scale=0.65]{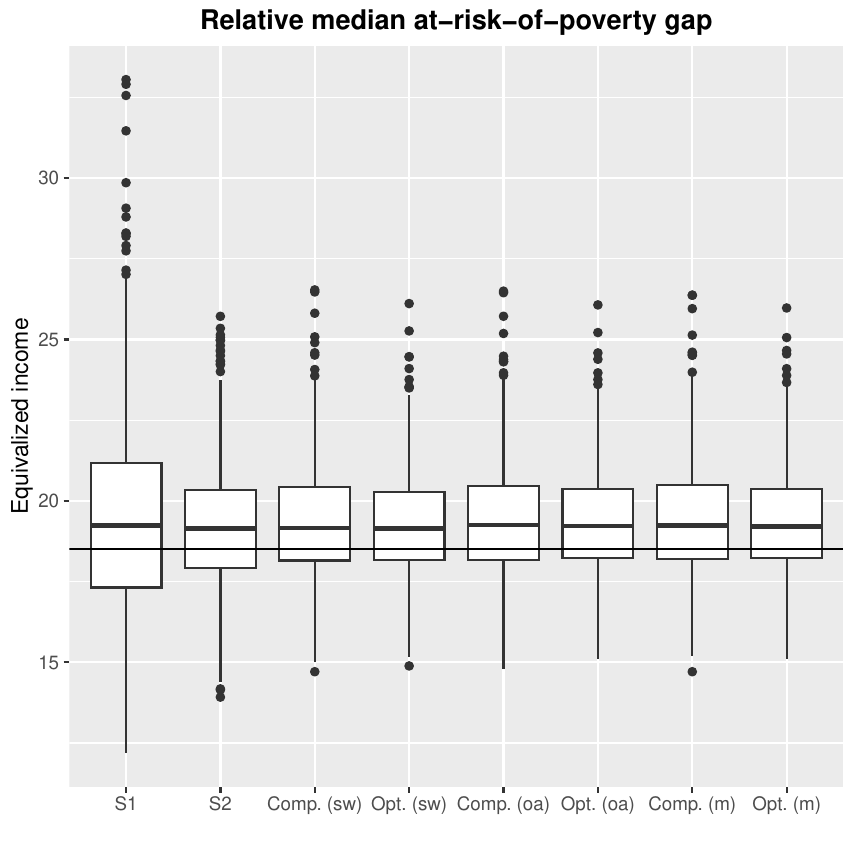}
\caption{\label{rmpg_500_1000} The RMPG of the equivalized income when $n_{h_1}=500$ and $n_{h_2}=1,000$. S1 and S2 stand for the first and the second sample, respectively. The weights are based on the sampling scheme (sw), aligning the RMPG (oa), and the alignment of the median (m). The estimators of the two samples are combined using a simple average (Comp.) and in an optimal way (Opt.). The horizontal line is the ``true" value on the population.}
\end{center}
\end{figure}

In Figures \ref{median_500_1000}-\ref{rmpg_500_1000}, we present the equivalized income when $n_{h_1}=500$ and $n_{h_2}=1,000$ for the median, the QSR, the Gini coefficient and the RMPG, respectively, under different scenarios about the weights used as well as the methods of combining the estimators of the two samples. The horizontal line is the ``true" value on the population from which the two samples are derived. Note that for the aligned estimates (the four boxplots at the right) the estimated quantity is the same for all the weights but for some weights we can see less variability due to the weight optimality. 

We need to mention that for the methods we use in order to combine the estimators of the two samples, we have only one estimate. Also, for these methods we have larger sample size since they make usage of both samples, and so the corresponding variances are smaller compared to the ones when we use just one sample (S1 and S2). According to Figures \ref{median_500_1000}-\ref{rmpg_500_1000}, the newly proposed approach appears to work satisfactorily, suggesting its usefulness in pooling information. Furthermore, as the proposed approach is based on the alignment method, it inherits all the favorable properties associated with alignment.

 In Table \ref{tabal} we provide the standard errors for the median, the QSR, the Gini coefficient and the RMPG for the equivalized income when $n_{h_1}=500$ and $n_{h_2}=1,000$ for the different scenarios about the weights used and the methods of combining the estimators of the two samples. We see that the standard errors of the median, the QSR and the Gini coefficient are less in the optimal way compared to when the two samples are combined using a simple average. Note that this holds for all the three scenarios about the weights used. The RMPG has the same behaviour among all the different scenarios about the weights used and the methods of combining the estimators of the two samples. Also, it is important to note that for the case of combining the two samples in the optimal way, the standard errors of the QSR, the Gini coefficient and the RMPG are slightly less when the sample weights used. However, under the optimal way of combining the two samples, the standard error of the median is less when its own alignment weights used.

\begin{table}[ht]
\centering
\begin{tabular}{rrrrrrr}
  \hline
& \multicolumn{2}{c}{Sample weights} &\multicolumn{2}{c}{Their own alignment} &\multicolumn{2}{c}{Alignment of median} \\ \hline
 Indicator & \begin{tabular}{@{}c@{}}Simple \\ average\end{tabular} & \begin{tabular}{@{}c@{}}Optimal \\ way\end{tabular} & \begin{tabular}{@{}c@{}}Simple \\ average\end{tabular} & \begin{tabular}{@{}c@{}}Optimal \\ way\end{tabular} & \begin{tabular}{@{}c@{}}Simple \\ average\end{tabular} & \begin{tabular}{@{}c@{}}Optimal \\ way\end{tabular} \\
  \hline
Median & 278.6186 & 261.1903 & 264.6534 & 246.4842 & 264.6534 & 246.4842\\
QSR  & 0.1192 & 0.1137 & 0.1194 & 0.1139 & 0.1192 & 0.1138 \\
Gini coefficient   & 0.5588 & 0.5327 & 0.5590 & 0.5329 & 0.5588 & 0.5327 \\
RMPG (\%) & 1.7184 & 1.5931 & 1.7167 & 1.5895 & 1.7187 & 1.5892 \\
   \hline
\end{tabular}
\caption{\label{tabal} The standard errors for the median, the QSR, the Gini coefficient and the RMPG for the equivalized income when $n_{h_1}=500$ and $n_{h_2}=1,000$, under the different scenarios about the weights used and the methods of combining the estimators of the two samples.}
\end{table}

\section{Concluding remarks}\label{concl}

In the present paper we extended the alignment method developed for totals to the case of complex statistics. The key ingredient is the use of the linearized variables of those complex statistics that allow to consider them in some linear and make the existing results applicable. We have also
shown based on some available data that the approach can
provide interesting gain in efficiency reducing the standard errors. 
While we demonstrated the approach based on social equality indicators, several other complex indicators can be benefited but the approach. 

Note that no special effort was given to make the computations efficient for working with large sample sizes. The calculation needs the second-order inclusion probabilities which for a sample of size $n$ is a $n\times n$ matrix. This can create memory and execution problems in practice. Alternatively one can take advantage of the block matrix creation of this matrix to speed up the estimation and save memory. We have implemented the first trivial approach.

Further checks with real data (in the sense of true and realistic sample sizes)
would be very useful as well as more complicated experiments where
a series of different estimators (indicators) of different
complexity and nature would have been estimated together.
Simulation results showed that the choice of some kind of optimal weights, while not  trivial, can benefit a lot the reduction of the variance. This topic needs to be considered further. 

\section*{Acknowledgments}

The second author participated back in 2016 in the project
"Integrated system of European Social Surveys – 
generalisation of the existing framework to cover 
longitudinal and other complex aspects"
funded by Eurostat under the contract N° 11111.2013.001-2014.526.
Discussions with Eurostat experts, especially with
Martin Karlberg and Fabrice Gras have helped a lot to shape the research
presented here.

\clearpage
\newpage
\bibliographystyle{apalike}
\bibliography{sample}

\pagebreak
\begin{center}
\textbf{\Large Supplementary Material}
\end{center}

\setcounter{figure}{0}
\setcounter{table}{0}
\makeatletter
\renewcommand{\thetable}{S\arabic{table}}
\renewcommand{\thefigure}{S\arabic{figure}}

In the Supplementary Material, we show the results for the second and the third scenarios about the sizes of the two samples, among which there is no overlapping. 

In the second scenario we have $n_{h_1}=n_{h_2}=1,000$, and in the third scenario we have $n_{h_1}=3,000$ and $n_{h_2}=1,000$.

\begin{figure}[h]
	\begin{center}
		\includegraphics[scale=0.7]{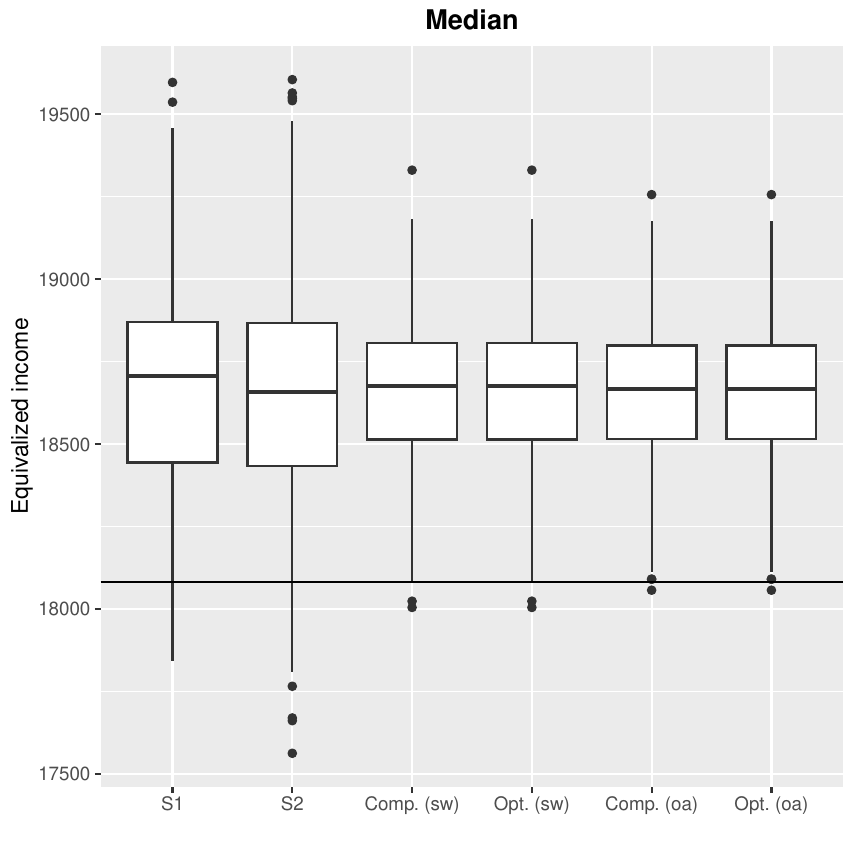}
		\caption{\label{median_1000_1000} The median of the equivalized income when $n_{h_1}=n_{h_2}=1,000$. S1 and S2 stand for the first and the second sample, respectively. The weights are based on the sampling scheme (sw) and aligning the median (oa). The estimators of the two samples are combined using a simple average (Comp.) and in an optimal way (Opt.). The horizontal line is the ``true" value on the population.}
	\end{center}
\end{figure}

\begin{figure}[H]
	\begin{center}
		\includegraphics[scale=0.7]{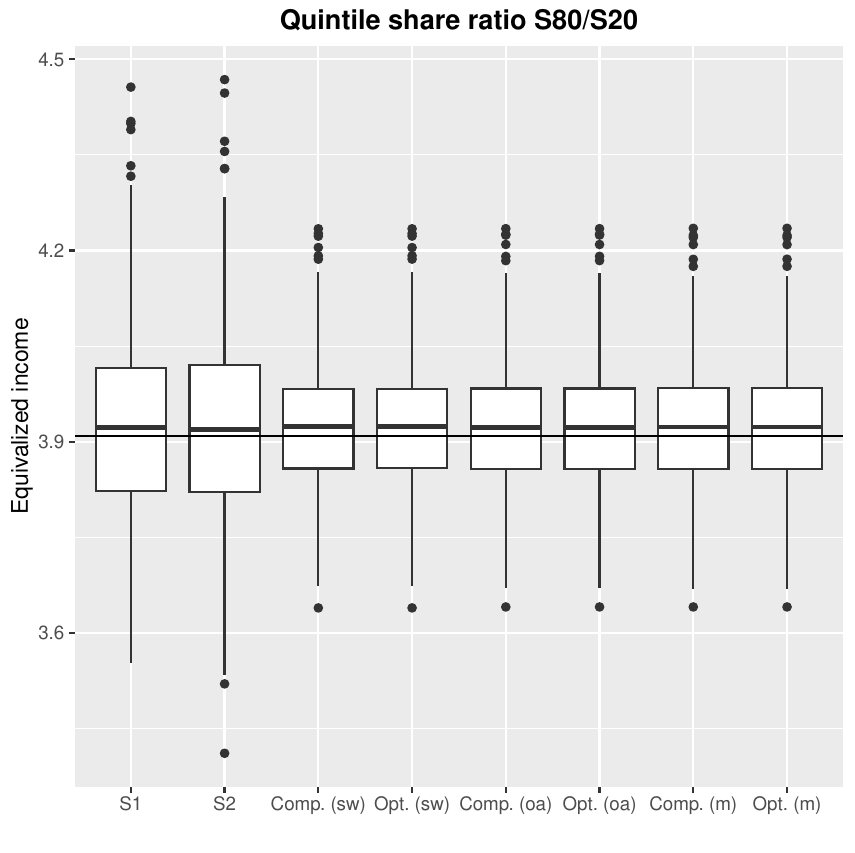}
		\caption{\label{qsr_1000_1000} The QSR of the equivalized income when $n_{h_1}=n_{h_2}=1,000$. S1 and S2 stand for the first and the second sample, respectively. The weights are based on the sampling scheme (sw), aligning the QSR (oa), and the alignment of the median (m). The estimators of the two samples are combined using a simple average (Comp.) and in an optimal way (Opt.). The horizontal line is the ``true" value on the population.}
	\end{center}
\end{figure}

\begin{figure}[H]
	\begin{center}
		\includegraphics[scale=0.7]{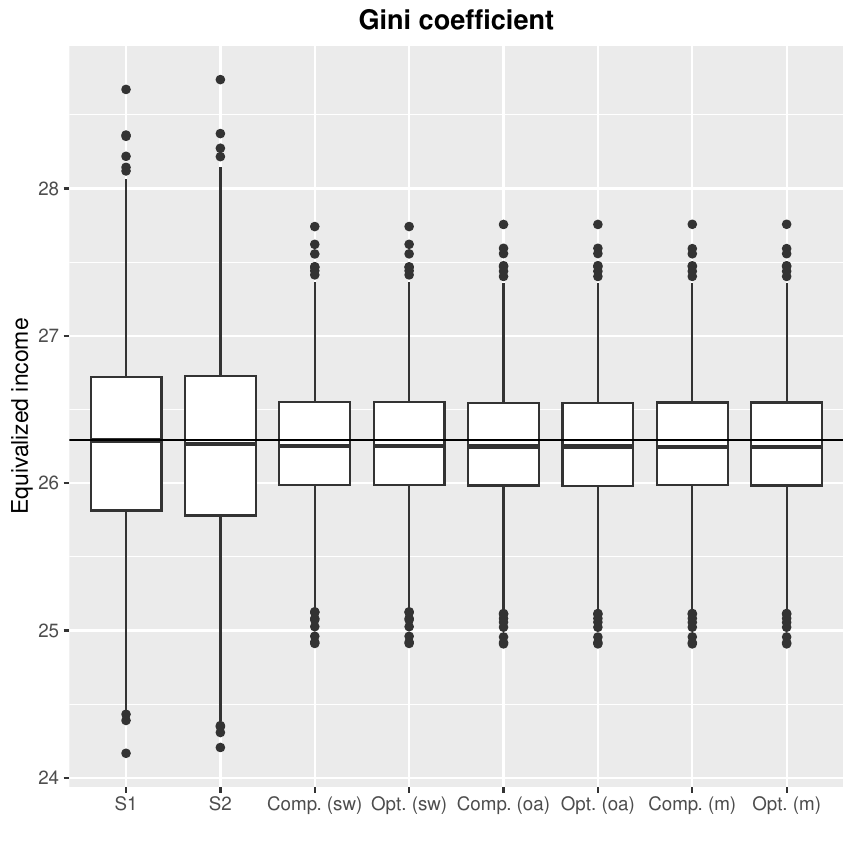}
		\caption{\label{gini_1000_1000} The Gini coefficient of the equivalized income when $n_{h_1}=n_{h_2}=1,000$. S1 and S2 stand for the first and the second sample, respectively. The weights are based on the sampling scheme (sw), aligning the Gini coefficient (oa), and the alignment of the median (m). The estimators of the two samples are combined using a simple average (Comp.) and in an optimal way (Opt.). The horizontal line is the ``true" value on the population.}
	\end{center}
\end{figure}

\begin{figure}[H]
	\begin{center}
		\includegraphics[scale=0.7]{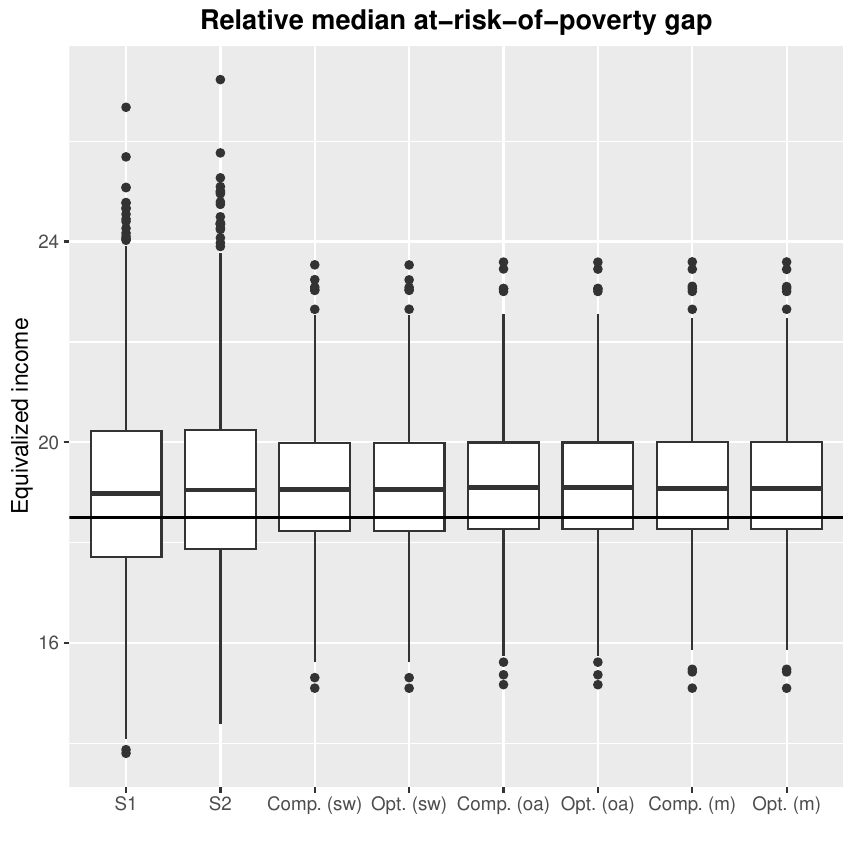}
		\caption{\label{rmpg_1000_1000} The RMPG of the equivalized income when $n_{h_1}=n_{h_2}=1,000$. S1 and S2 stand for the first and the second sample, respectively. The weights are based on the sampling scheme (sw), aligning the RMPG (oa), and the alignment of the median (m). The estimators of the two samples are combined using a simple average (Comp.) and in an optimal way (Opt.). The horizontal line is the ``true" value on the population.}
	\end{center}
\end{figure}

In Figures \ref{median_1000_1000}-\ref{rmpg_1000_1000}, we present the equivalized income when $n_{h_1}=n_{h_2}=1,000$ for the median, the QSR, the Gini coefficient and the RMPG, respectively, under different scenarios about the weights used as well as the methods of combining the estimators of the two samples. The horizontal line is the ``true" value on the population from which the twp samples are derived.

\newpage

\newpage

\newpage

In Table \ref{tabal_1000_1000} we provide the standard errors for the median, the QSR, the Gini coefficient and the RMPG for the equivalized income when $n_{h_1}=n_{h_2}=1,000$ for the different scenarios about the weights used and the methods of combining the estimators of the two samples.

\begin{table}[H]
	\centering
	\begin{tabular}{rrrrrrr}
		\hline
		& \multicolumn{2}{c}{Sample weights} &\multicolumn{2}{c}{Their own alignment} &\multicolumn{2}{c}{Alignment of median} \\ \hline
		Indicator & \begin{tabular}{@{}c@{}}Simple \\ average\end{tabular} & \begin{tabular}{@{}c@{}}Optimal \\ way\end{tabular} & \begin{tabular}{@{}c@{}}Simple \\ average\end{tabular} & \begin{tabular}{@{}c@{}}Optimal \\ way\end{tabular} & \begin{tabular}{@{}c@{}}Simple \\ average\end{tabular} & \begin{tabular}{@{}c@{}}Optimal \\ way\end{tabular} \\
		\hline
		Median & 208.0224 & 208.0218 & 198.8549 & 198.8531 & 198.8549 & 198.8531\\
		QSR  & 0.0932 & 0.0932 & 0.0933 & 0.0933 & 0.0934 & 0.0934 \\
		Gini coefficient   & 0.4381 & 0.4381 & 0.4388 & 0.4388 & 0.4388 & 0.4388 \\
		RMPG (\%) & 1.2672 & 1.2672 & 1.2728 & 1.2729 & 1.2699 & 1.2700 \\
		\hline
	\end{tabular}
	\caption{\label{tabal_1000_1000} The standard errors for the median, the QSR, the Gini coefficient and the RMPG for the equivalized income when $n_{h_1}=n_{h_2}=1,000$, under the different scenarios about the weights used and the methods of combining the estimators of the two samples.}
\end{table}

%%%%%%%%%%%

\begin{figure}[h]
	\begin{center}
		\includegraphics[scale=0.7]{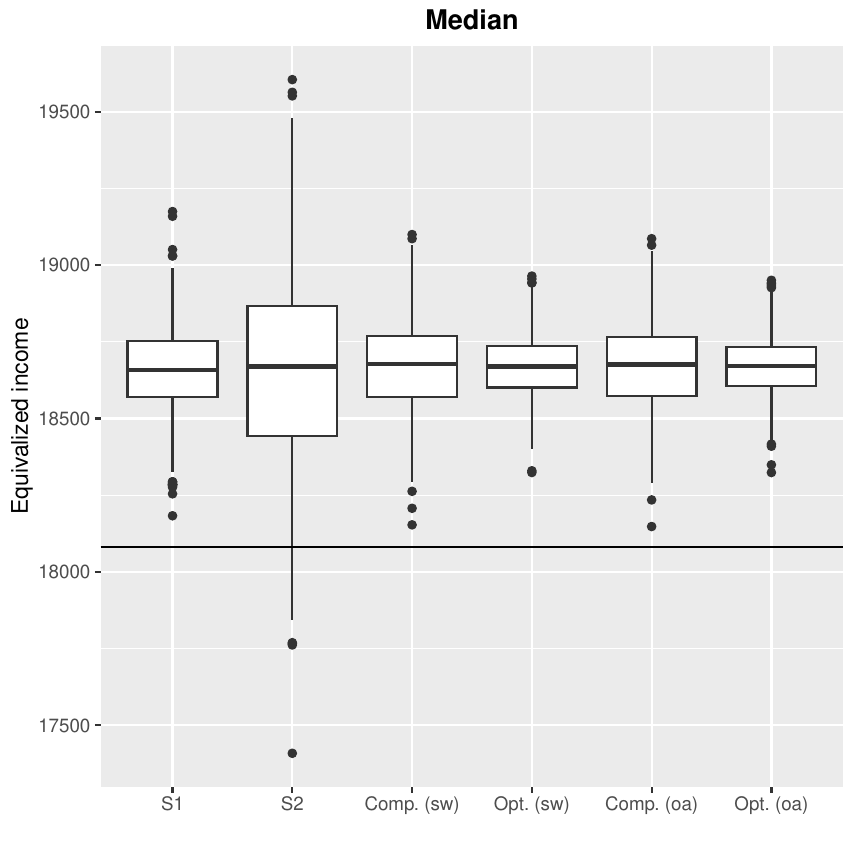}
		\caption{\label{median_3000_1000} The median of the equivalized income when $n_{h_1}=3,000$ and $n_{h_2}=1,000$. S1 and S2 stand for the first and the second sample, respectively. The weights are based on the sampling scheme (sw) and aligning the median (oa). The estimators of the two samples are combined using a simple average (Comp.) and in an optimal way (Opt.). The horizontal line is the ``true" value on the population.}
	\end{center}
\end{figure}

\begin{figure}[H]
	\begin{center}
		\includegraphics[scale=0.7]{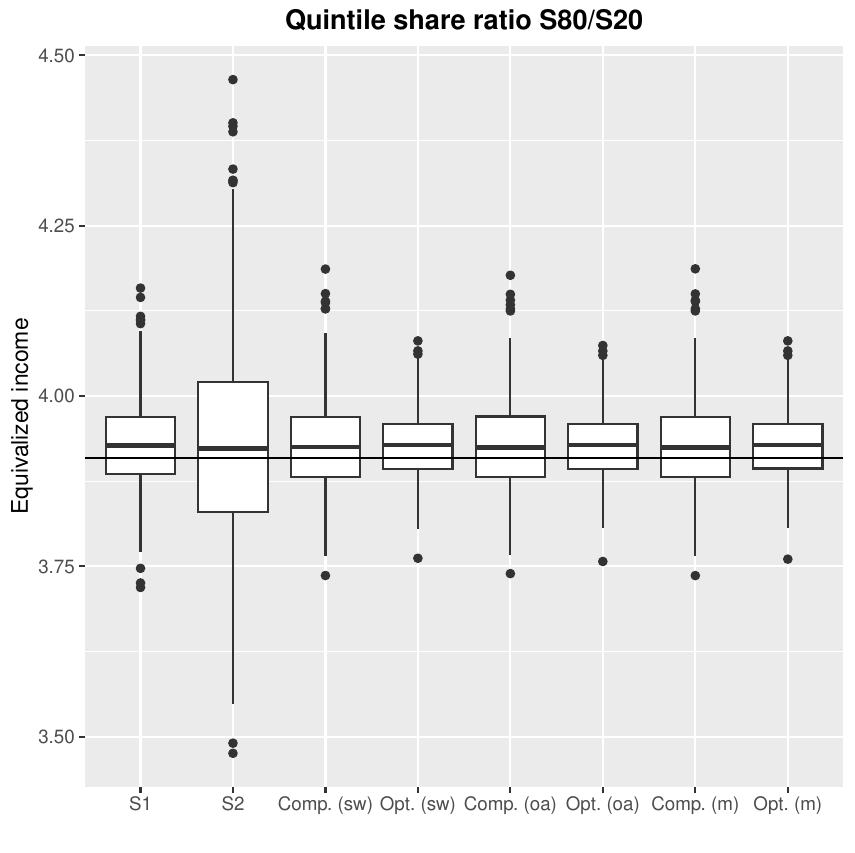}
		\caption{\label{qsr_3000_1000} The QSR of the equivalized income when $n_{h_1}=3,000$ and $n_{h_2}=1,000$. S1 and S2 stand for the first and the second sample, respectively. The weights are based on the sampling scheme (sw), aligning the QSR (oa), and the alignment of the median (m). The estimators of the two samples are combined using a simple average (Comp.) and in an optimal way (Opt.). The horizontal line is the ``true" value on the population.}
	\end{center}
\end{figure}

\begin{figure}[H]
	\begin{center}
		\includegraphics[scale=0.7]{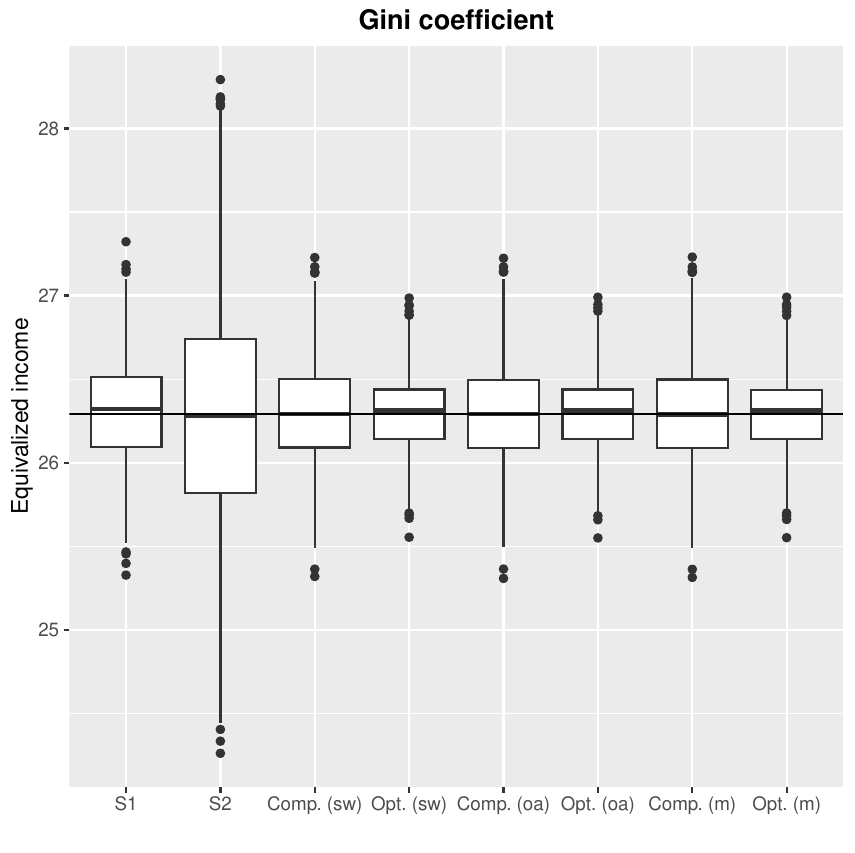}
		\caption{\label{gini_3000_1000} The Gini coefficient of the equivalized income when $n_{h_1}=3,000$ and $n_{h_2}=1,000$. S1 and S2 stand for the first and the second sample, respectively. The weights are based on the sampling scheme (sw), aligning the Gini coefficient (oa), and the alignment of the median (m). The estimators of the two samples are combined using a simple average (Comp.) and in an optimal way (Opt.). The horizontal line is the ``true" value on the population.}
	\end{center}
\end{figure}

\begin{figure}[H]
	\begin{center}
		\includegraphics[scale=0.7]{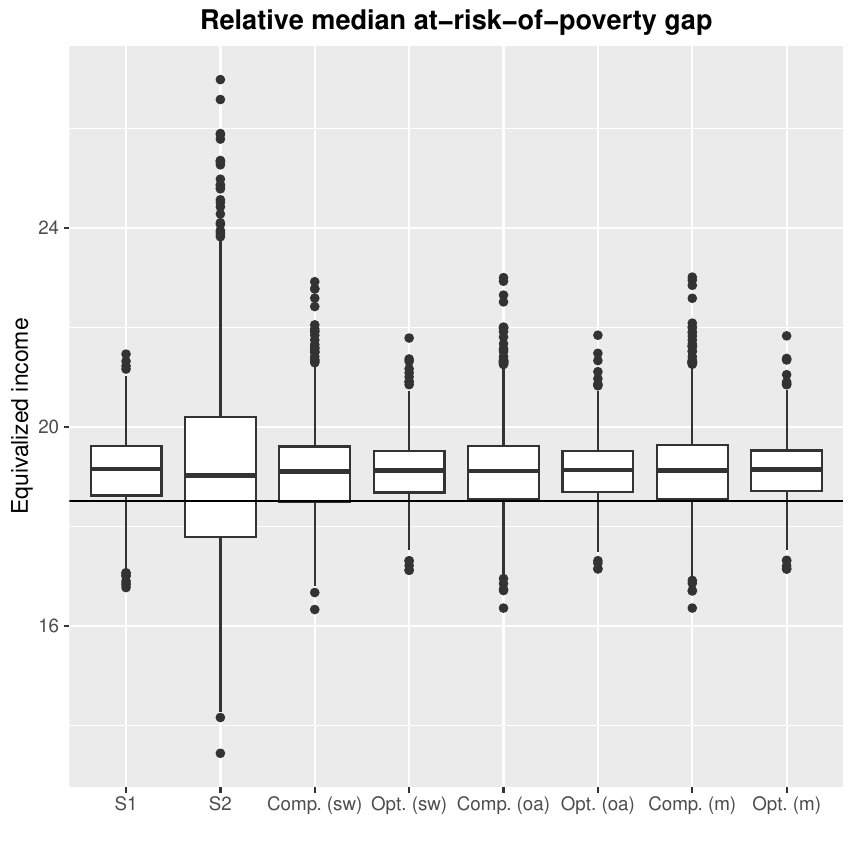}
		\caption{\label{rmpg_3000_1000} The RMPG of the equivalized income when $n_{h_1}=3,000$ and $n_{h_2}=1,000$. S1 and S2 stand for the first and the second sample, respectively. The weights are based on the sampling scheme (sw), aligning the RMPG (oa), and the alignment of the median (m). The estimators of the two samples are combined using a simple average (Comp.) and in an optimal way (Opt.). The horizontal line is the ``true" value on the population.}
	\end{center}
\end{figure}

In Figures \ref{median_3000_1000}-\ref{rmpg_3000_1000}, we present the equivalized income when $n_{h_1}=3,000$ and $n_{h_2}=1,000$ for the median, the QSR, the Gini coefficient and the RMPG, respectively, under different scenarios about the weights used as well as the methods of combining the estimators of the two samples. The horizontal line is the ``true" value on the population from which the twp samples are derived.

\newpage

\newpage

\newpage

In Table \ref{tabal_3000_1000} we provide the standard errors for the median, the QSR, the Gini coefficient and the RMPG for the equivalized income when $n_{h_1}=3,000$ and $n_{h_2}=1,000$ for the different scenarios about the weights used and the methods of combining the estimators of the two samples.

\begin{table}[H]
	\centering
	\begin{tabular}{rrrrrrr}
		\hline
		& \multicolumn{2}{c}{Sample weights} &\multicolumn{2}{c}{Their own alignment} &\multicolumn{2}{c}{Alignment of median} \\ \hline
		Indicator & \begin{tabular}{@{}c@{}}Simple \\ average\end{tabular} & \begin{tabular}{@{}c@{}}Optimal \\ way\end{tabular} & \begin{tabular}{@{}c@{}}Simple \\ average\end{tabular} & \begin{tabular}{@{}c@{}}Optimal \\ way\end{tabular} & \begin{tabular}{@{}c@{}}Simple \\ average\end{tabular} & \begin{tabular}{@{}c@{}}Optimal \\ way\end{tabular} \\
		\hline
		Median & 140.4034 & 103.2093 & 137.0195 & 98.2817 & 137.0195 & 98.2817\\
		QSR  & 0.0662 & 0.0480 & 0.0659 & 0.0478 & 0.0660 & 0.0478 \\
		Gini coefficient   & 0.3116 & 0.2235 & 0.3116 & 0.2235 & 0.3119 & 0.2236 \\
		RMPG (\%) & 0.9190 & 0.6425 & 0.9092 & 0.6347 & 0.9211 & 0.6415 \\
		\hline
	\end{tabular}
	\caption{\label{tabal_3000_1000} The standard errors for the median, the QSR, the Gini coefficient and the RMPG for the equivalized income when $n_{h_1}=3,000$ and $n_{h_2}=1,000$, under the different scenarios about the weights used and the methods of combining the estimators of the two samples.}
\end{table}

\end{document}